\documentstyle[10pt,aas2pp4]{article}
\newcommand{\etal}{{\it et al. }}
\def\cm2{\,{\rm cm}^2}
\begin{document}

\title{A Broad Band and Large Area X-Ray Omni Sky Monitor (BLOSM)~\footnote{
       Contributed paper to ``All-Sky X-Ray Observations in the Next Decade: A
       Workshop for ASM and GRB Missions in the X-Ray Band, The Institute
       of Physical and Chemical Research (RIKEN), Wako, Saitama, Japan,
       March 3-5, 1997.} }
\author{ W. Zhang, R. Petre, A. N. Peele$^1$, K. Jahoda, 
	 F. E. Marshall, Y. Soong$^2$, and N. E. White}  
\affil{Laboratory for High Energy Astrophysics \\ 
       Goddard Space Flight Center\\
       Greenbelt, Greenbelt, MD 20771\\
      $^1$ National Research Council Research Associate \\
      $^2$ $also$ Universities Space Research Association\\
       {\it E-mail: William.W.Zhang.1@gsfc.nasa.gov} } 

\begin{abstract}
We present a conceptual design for a new X-ray all sky
monitor (ASM). Compared with previous ASMs, its salient features are: 
(1) it has a focusing capability that increases the signal to
background ratio by a factor of 3; (2) it has a broad-band width:
200 eV  to 15 keV; (3) it has a large X-ray collection area:
$\sim 10^2\cm2$; (4) it has a duty cycle of nearly 100\%, and (5) it
can measure the position of a new source with an accuracy of a few
minutes of arc. These features combined open up an opportunity for
discovering new phenomena as well as monitoring existing
phenomena with unprecedented coverage and sensitivity.
\end{abstract}

\keywords{Instrumentation: X-ray optics, all sky monitor}

\section{Introduction}
ASMs have played an important role in the development of X-ray
astronomy. A large number of transients  have been discovered and a
large number of bright persistent sources have been monitored with
them.  Their role as watchdog to alert pointing instruments has been
most prominent.

Previous all sky monitors suffer from several common shortcomings.
First, they are only sensitive to X-rays above 2 keV. Second, they have
relatively small effective areas or very small time-averaged effective
areas. Third, they do not have any focusing capability. Combining
all these factors, their measurement limit is, at best, of the order
of several mCrab in several hours or in one day.

In this paper we further develop a concept originally put forth by
Schmitt in 1975. With advances in both X-ray optics and detector
technology in the past two decades, this concept now is feasible.
We will present 
simulations to show the salient features of such an ASM. It can either
fly as a small free-flyer or as one of instruments on a large satellite
as previous ASMs.

Table~\ref{zand.table}
summarizes the most important characteristics of past ASMs.
The last row shows the expected capabilities of BLOSM. It
represents a significant step in improving the capability and
sensitivity of X-ray ASM.

\section{Instrument and Mission Concept}

The large field of view (FOV) optic proposed by Schmitt in 1975 is 
focusing in one dimension (see Peele \etal\ 1997 for an illustration of
this optic) by flat reflectors mounted along the radial directions of
a cylinder. With one module, it focuses a point source in its FOV to a
line on the focal surface. In our conceptual design here, we will have
two modules situated in perpendicular orientations so that the two
coordinates of a source can be obtained simultaneously.

\begin{figure*}
\vskip 3.5 in
\includegraphics{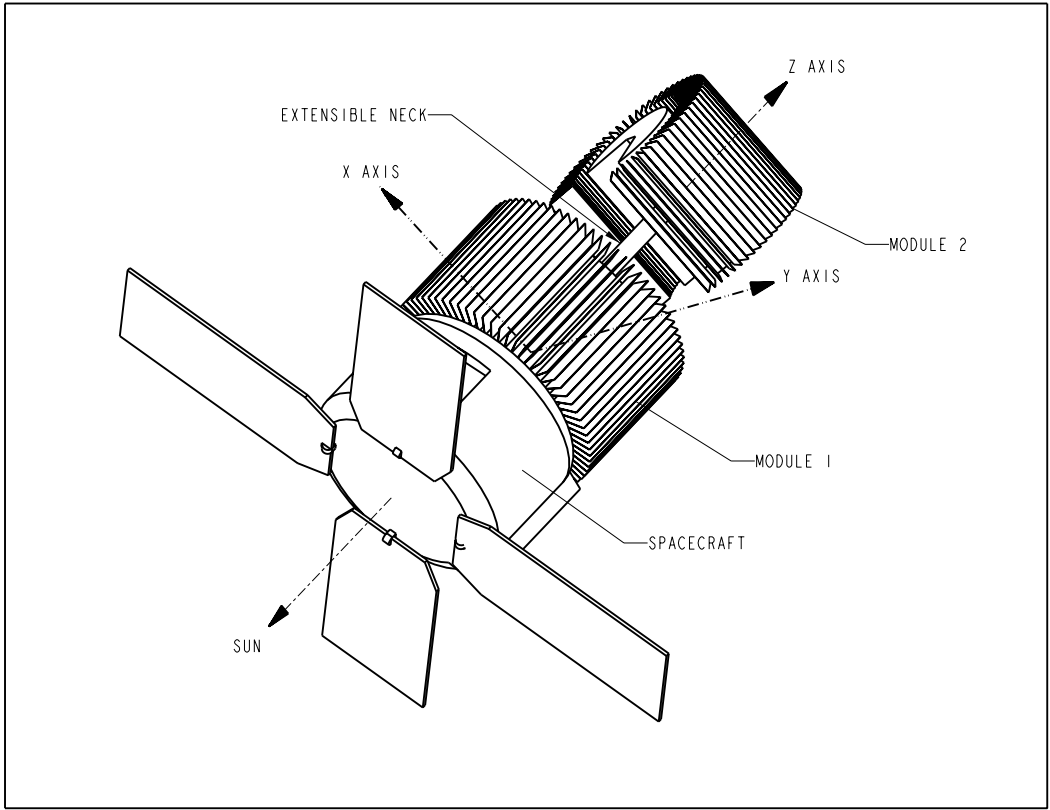}
\caption{Conceptual configuration of a free flyer that carries
	  the all sky monitor described in this paper. With the
	  parameters in this paper, the entire satellite here fits
	  in the latest NASA Small Explorer envelope. The extensible neck 
	  is meant to be compressed at launch and extended 
	  after launch.}
\label{lobster.ps}
\end{figure*}

As shown in Figure~\ref{lobster.ps}, the instrument consists of two
modules.  Module 1 is a cylinder 100 cm in diameter at its outer edge
and 50 cm in height. Its focal length is 25 cm. Module 2 is 83\% (i.e.,
300 out of 360 degrees) of a cylinder 70 cm (i.e., $100/\sqrt{2}$) in
diameter and 70 cm in height. Its focal length is 17.5 cm. Module 2
sits about 50 cm away on top of module 1. The 50 cm distance is to
ensure that the vanes at the lower side of module 2 are not blocked by
module 1.

For the purpose of discussion in this paper, without loss of
generality, we will adopt a fake equatorial coordinate system in which
the $x$-axis points to direction ($\alpha,\delta)=(0^\circ,0^\circ)$, and
$y$-axis $(90^\circ,0^\circ)$, and $z$-axis $(0^\circ, 90^\circ)$, and
the sun is in position ($0^{\circ}$, $-90^{\circ}$), as shown in 
Figure~\ref{lobster.ps}.
The cylindrical surface of module 1 is defined by: $x^2+y^2=50^2$
and $-25 < z < 25$, and module 2 by: $y^2 + (z-135)^2=35^2$ and $ -35 < x
< 35,$ where all numbers are in cm.  The entire satellite
spins around the direction to the sun, which, in the specific
coordinate system, corresponds to the $z$-axis.

\begin{figure*}
\vskip 3.5 in
\includegraphics{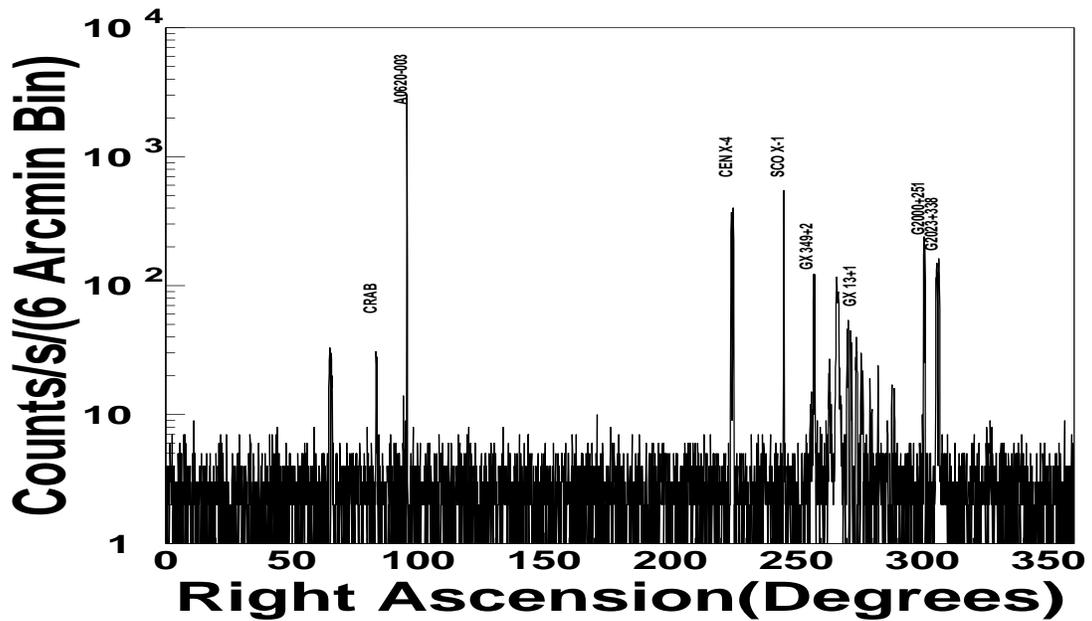}
\caption{A snap shot of the sky with one second of observation.
	 In the fake equatorial coordinate system adopted here,
	 the right ascension coincides with the azimuthal angle
	 along the circumference of module 1.}
\label{mod01_1s.ps}
\end{figure*}

The cylindrical position-sensitive detectors need only to measure
position along the circumferential direction. Every time interval,
each model produces a map similar to  Figure~\ref{mod01_1s.ps}.

\section{Source Mapping Deconvolution}

For the purpose of illustration, we need to define two angular
variables which correspond to the position of a source on the
focal surfaces. For module 1, the angle $\theta_1$ measures
from the positive $x$-axis with $y$-axis at $90^\circ$, i.e.,
$\theta_1$ is the azimuthal angle of module 1. 
For module 2, the angle $\theta_2$ measures from the $y$-axis
with the $z$-axis at $90^\circ$, i.e., $\theta_2$ is the
azimuthal angle of module 2. These two variables are fixed
with respect to the modules.

At time $t=0$ for a source at $(\alpha, \delta)$, it has the following
angles in the two modules:

\begin{eqnarray}
 module 1:  \theta_1 = \alpha, &   \\
 module 2:  \theta_2 = arctan(\cos\delta\sin\alpha, \sin\delta), &
\end{eqnarray}
which should be interpreted as $\sin\theta_2 = \cos\delta\sin\alpha$
and $\cos\theta_s =\sin\delta$.
For example, two sources with identical $\alpha$, but different
$\delta$, will have one image in module 1, but two separate images in
module 2. A new source and its two coordinates can be easily identified
this way.  The rotation of the satellite can be taken into account by
replacing $\alpha$ in the above equations with $\alpha - \Omega t$,
where $\Omega$ is the angular velocity of the satellite, and $t$ time.

\section{Angular Resolution, Effective Area, and Background}

\begin{figure*}
\vskip 3.5 in
\includegraphics{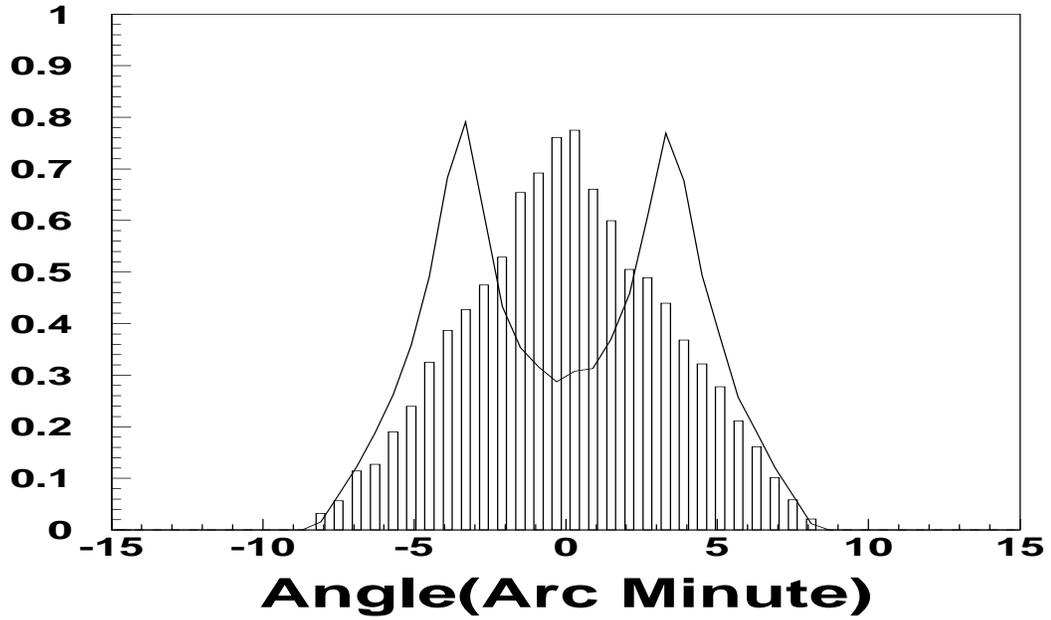}
\caption{Comparison of images achieved with reflectors with no
	 thickness and with thickness of 0.05 cm which is 
	 realistically achievable. In both cases, the linear
	 distance between two adjacent reflectors is 0.123 cm and 
	 the radius of the cylinder 50 cm.}
\label{psf.ps}
\end{figure*}

In the ideal case where reflectors are infinetly thin and they are 
mounted perfectly along their local radial directions, the angular
resolution of the system is the same as the angle between two adjacent
reflectors. Figure~\ref{psf.ps}
compares the ideal situation and a realistic
situation where the reflector thickness is 0.05 cm. In both cases
the angular spacing and other dimensions are the same.

\begin{figure*}
\vskip 3.5 in
\includegraphics{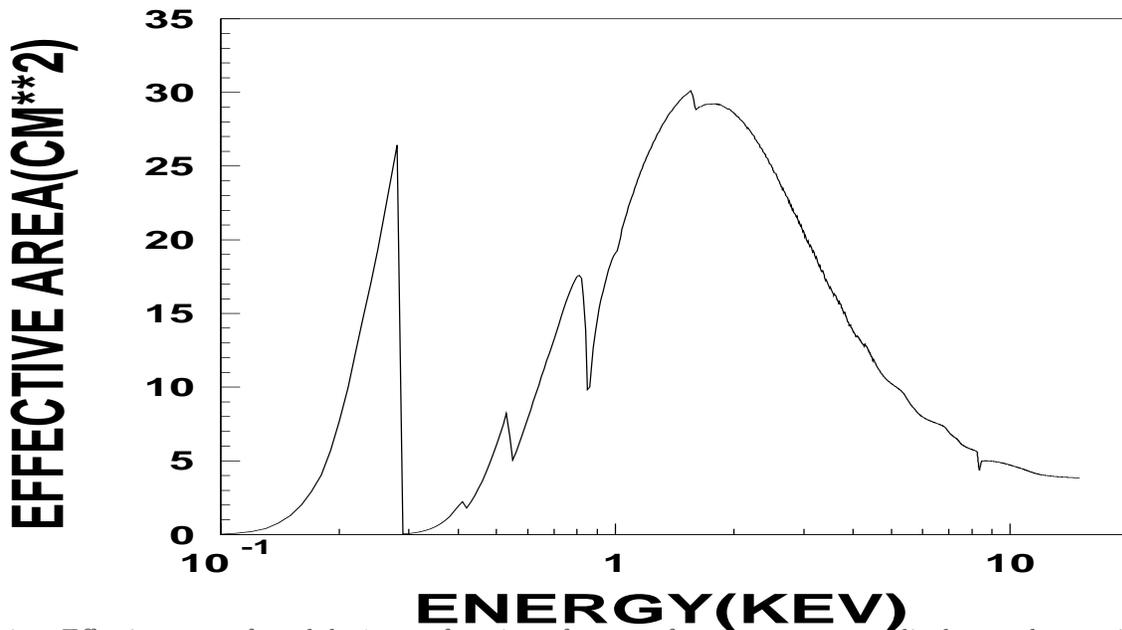}
\caption{Effective area of module 1 as a function of energy for a source
	 perpendicular to the $z$-axis. We have assumed the window is
	 the same as in Priedhorsky, Peele, and Nugent (1996).}
\label{aeff.ps}
\end{figure*}

Assuming that the mirrors are coated with Ni with perfect smoothness
and the length of each reflector along the radial direction is 3.3 cm
(see Peele \etal\ 1997) for module 1, Figure~\ref{aeff.ps}
shows the effective area
of module 1 as a function of energy for a source perpendicular to the
$z$-axis.  Module 2 has exactly the same effective area for a source
perpendicular to the $z$-axis. Though its radius is smaller by a factor
of $\sqrt{2}$, it is longer by the same factor.

There are three sources of background: (1) detector internal (non-Xray)
background,
(2) background contribution from sources in the same annulus of the
sky that share the same spot on the focal surface, and (3) the
diffuse X-ray background. For the system as outlined in this
paper, the detector internal background is  much lower
than the other two, thus we will ignore it for now. The background
contributions of other sources, equivalent to source confusion, 
though complicated, should not be a problem for relatively bright sources. 
For the purpose of this paper, we will only consider the 
diffuse X-ray background.

\begin{figure*}
\vskip 3.5 in
\includegraphics{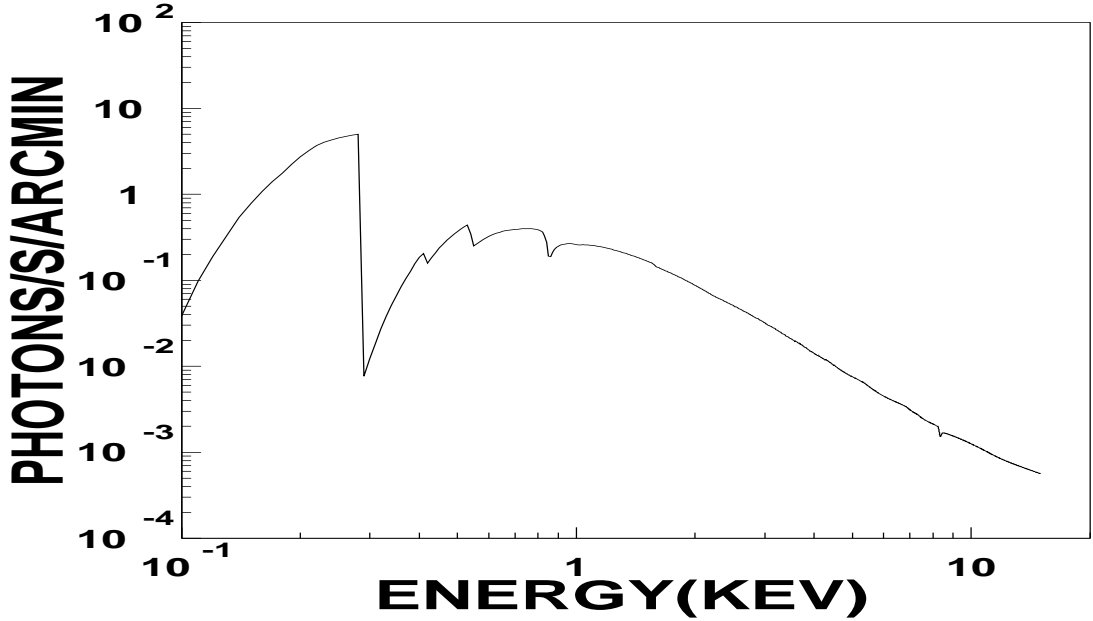}
\caption{Background event rate as a function of energy for module 1.
         Note here the unit of the ordinate is $photons/s/ArcMin$.
	 For the optimal parameters used in this paper, one should 
	 use a 15 ArcMin angular size to estimate background.}
 \label{bkgnd.ps}
\end{figure*}

The diffuse X-ray background flux can be parameterized (Priedhorsky \etal\ 1996)
as

\begin{eqnarray}
    f = & 20E^{-1.89}   & 0.05 \le E \le 0.44\  keV,  \\
	& 16E^{-2.16}   & 0.44 \le E \le 2.23\  keV,  \\
	& 9.2E^{-1.47}  & 2.23 \ keV \le E,
\end{eqnarray}
where $E$ is in $keV$ and $f$ in $photons/cm^2/s/sr$. The result of
simulations with this flux is shown in Figure~\ref{bkgnd.ps}.

\section{Practical Considerations}

We have been systematically investigating different materials to
characterize their suitability as reflectors. We have studied Mylar,
Kapton, and PlexiGlas and have found that those materials, even though
they can be made very thin and strong, do not satisfy flatness
requirements. We have found that thin glass (0.05 cm) which is
commercially produced for flat-panel LCD displays has the desired
flatness. We have optically tested a few pieces and found that, even
though not all of them satisfy the flatness requirements, a significant
fraction of them have flatness better that one minute of arc. We are
continuing this investigation to identify the best material. For the
purpose of this paper, we will assume that we use glass reflectors with
a thickness of 0.05 cm. The angular spacing between two adjacent
reflectors is assumed to be 10 minutes of arc, which corresponds to a
linear distance of 0.12 and 0.08 cm for modules 1 and 2, respectively.
These linear spacings are quite comfortable for mechanically mounting
those reflectors to a structure.

In general, these glass sheets are quite smooth. Coated with 
either gold or nickel, their measured X-ray reflectivity is nearly identical
to theoretical expectations.

A challenging part of the building the system as outlined in this 
paper is the detectors. We require two detectors each with an active
area on the order of 7,000 cm$^2$. They must have thin window 
and good one-dimensional position resolution. We have been
investigating the possibility of using the micro-strip proportional
counter
technology. The glass plate will have anode and cathode traces
laid on them photolithographically with a pitch of, say, 
300 $\mu$m. Each trace is read out by its own analog electronics
to achieve maximum position resolution. Since we do not require
position resolution along the longitudinal directions of 
the cylinders, the readout electronics can be made simple.

With all these parameters, Table~\ref{mass.table}
list the key components of
this system and their estimated masses.

\section{Scientific Merits}

\begin{figure*}
\vskip 3.5 in
\includegraphics{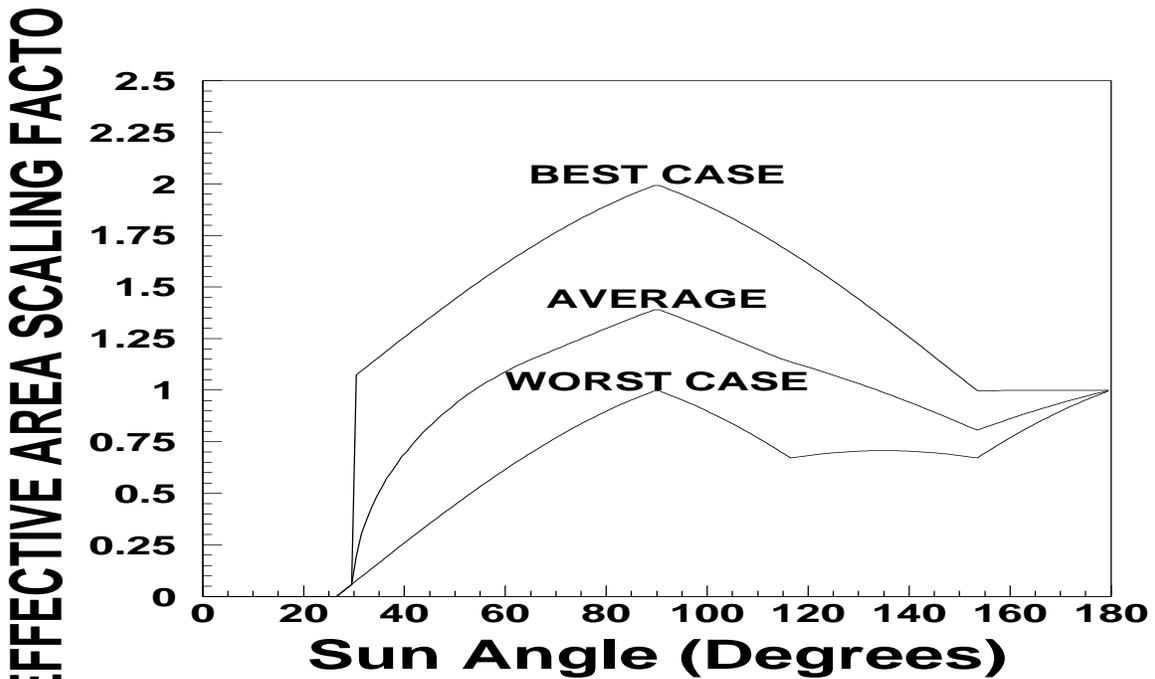}
\caption{Effective area scaling factor as a function of sun angle.
	 The effective area of BLOSM at a given energy for
	 a source is the product of this factor with the corresponding
	 area in Figure~\ref{aeff.ps}.}
 \label{aeff_factor.ps}
\end{figure*}

Like previous ASMs, BLOSM is capable of monitoring long-term
behavior of bright sources. But the most significant aspect 
of BLOSM is that it covers the soft band below 2 keV and
its coverage of every part of the sky except for sources within
$30^\circ$ of the sun. With these two features, we expect 
it to discover many fast (lasting on the order of seconds or
minutes) and/or soft transients. In this section, we give 
quantitative estimates of BLOSM count rates for several known
phenomena. Note that in doing these estimates, we will assume 
the source is `on-axis,' i.e., it is in a direction perpendicular
to the symmetry axis of module 1. The count rates are for module 1
only. For a source that is not on-axis, it may get more or less counts
than these estimates, depending on its position. Figure~\ref{aeff_factor.ps}
shows
how the effective area varies as a function of sun angle, where a
factor of 1 corresponds to the estimate here.

\begin{enumerate}
  \item As a point of reference, we have included the expected 
  count rates from the Crab nebular/pulsar. We have used a
  power law spectrum with a photon index of -2.05 and an
  $n_H=3.0\times 10^{21} cm^{-2}$.

  \item Gamma Ray Bursts: Since their discovery nearly three decades ago,
  gamma ray bursts have been the most enigmatic astrophysical phenomenon.
  It has been generally agreed upon in the last few years that one
  of potentially very useful measurements is to measure the
  galactic absorption at low energies. For the estimates in
  Table~\ref{rate.table}, we have used the flux measured with the Ginga 
  gamma ray burst detector (T.E. Strohmayer, personal communication) which
  corresponds to the spectral characterization of Band \etal (1993)
  with $\alpha=-0.9, \beta=-5.0, E_0=100 keV,$ and $A=0.5$. We have considered
  two cases. In the first case, we extrapolate this spectrum all
  the way down to 0.1 keV with no interstellar absorption.
  In the second case, we have assumed an interstellar absorption
  corresponding to $n_H=5\times 10^{22} cm^{-2}$ and the
  Morrison and McCammon (1983) cross sections. It is clear that 
  BLOSM is well suited for differentiating the galactic and
  extra-galactic origin theories of gamma ray bursts.

  \item BLOSM is perhaps the first ASM capable of systematic
  detection of X-ray bursts from galactic sources. Of the
  120 or so cataloged galactic low mass X-ray binaries, only 
  40 or so have been observed to emit Type-I bursts. With a few
  years of operation, BLOSM should be able to detect X-ray bursts
  from most of these sources. On the other hand, if no Type-I
  bursts are detected from a significant number of those sources,
  it may indicate that some of them may very well be black
  hole systems. For the estimates in Table~\ref{rate.table}, we have assumed
  that the neutron star has a radius of 10 km and is at a distance of
  10 kpc. It is clear that BLOSM can detect most of Type-I 
  X-ray bursts. With its position measurement accuracy of 0.1
  degrees, BLOSM can associate detected bursts with their
  persistent counterparts.

  \item Active Galactic Nuclei:  Long-term variability of AGNs
  has been the subject of intense organized campaigns in recent
  years. In Table~\ref{rate.table} we show the BLOSM count rate for a 1 mCrab
  AGN. The spectrum we have used is a power law with a photon
  index of -1.7 with $n_H=3.0\times 10^{20} cm^{-2}$ (Mushotzky \etal\ 1993).
  It takes about 50,000 seconds of observation to detect a
  1 mCrab AGN at 10$\sigma$ level. Therefore BLOSM probably can
  monitor a few bright AGNs on a daily basis.

  \item WGA Catalog/ROSAT Transients: A large number of soft transients 
  have been detected by ROSAT during its observations in the last five
  years (White 1997 and Angelini, Giommi, \& White 1996). With its 
  all sky coverage, BLOSM is expected to detect the brighter ones 
  and monitor them on a daily basis. For example, for a source
  with a flux of $1.6 \times 10^{-11} ergs/s/cm^2$ at the detector in
  the band of 0.5-1.5 keV, BLOSM will detect 0.1 counts/s over 
  a background of 4 counts/s. 

\end{enumerate}

\section{Conclusion}

In summary, we have demonstrated that a one-dimensional focusing all
sky monitor as outlined in this paper stands a significant step forward
in the direction of larger area and true all sky coverage. Compared
with previously-flown and currently flying ASMs, its improvement in
source location precision and monitoring sensitivity is well over an
order of magnitude. In addition to the capabilities of traditional
X-ray all sky monitors, it is capable of detecting gamma ray bursts,
soft transients, X-ray bursts, and monitoring a number of AGNs on a daily
basis. We conclude by pointing out that the instrument as outlined
here is suitable to fly either as a free flyer, such as a small
explorer, or as one of many instruments on a larger satellite.

\section*{Acknowledgment} We would like to thank Scott Murphy 
for help in preparing this paper.

\section*{References}

   L. Angelini, P. Giommi, and N.E. White, 1996, in {\em Rontgendtrahlung
    from the Universe}, ed. H.U. Zimmermann, J.E. Trumper, and H. York,
    pp. 645-646.

   D. Band \etal, 1993, ApJ, Vol 413, p. 281.

   R. Morrison and D. McCammon, 1983, ApJ. Vol. 270, pp.119-112.

   R.F. Mushotzky, C. Done, and K.A. Pounds, 1993, in {\em Annu. Rev.
    Astron. Astrophys.}, Vol. 31, pp. 717-761.

   A. Peele \etal\ 1997, in these proceedings.

   W.C. Priedhorsky, A.G. Peele, and K.A. Nugent, 1996, MNRAS, Vol. 279,
    pp. 733-750.

   W.K.H. Schmidt 1975, Nucl. Instr. Methods, Vol. 127, pp. 285-292.

   N.E. White 1997, in these proceedings.

   J.J.M. in t Zand \etal\ 1994, SPIE Proceedings, Vol. 2279,
    pp. 458-468.

\pagebreak

\pagebreak

\pagebreak

\begin{table*}
\caption{Key parameters of past, present, and future X-ray all sky
	 monitors. Adapted with modifications from in t Zand \etal
	 For MOXE and BLOSM, both of which have true all sky coverage,
	 their sensitivities correspond to 1.5 hours observations.
	 For others, the sensitivity column corresponds to the time
	 required to cover 75\% of the sky.}

\begin{center}
\begin{tabular}{lccccc} 
\hline\hline
Instrument & Operational & Instantaneous & Angular & Bandpass & Sensitivity  \\
	   & Period      & Sky Coverage & Resolution ($^\circ$)& (keV)  & (mCrab)     \\
\hline
Vela 5B XC & 1969-1979   & $5\times 10^{-4}$& 6.1 & 3-12   & 400         \\
Ariel V ASM& 1974-1980   & $1\times 10^{-2}$& 10  & 3-6    & 170          \\
Ginga ASM  & 1987-1991   & $5\times 10^{-4}$& 0.2 & 2-20   &  50       \\
Watch/Granat& 1989-      & $ 1 $            & 2   & 6-180  & 100       \\
BATSE/CGRO &  1990-      & $ 1 $            & 6   & 20-100 & 60        \\
ASM/RXTE   &  1996-      & $ 5\times 10^{-2}$& 0.2 & 2-10   & 20        \\
MOXE/SXG   &  1997-      & $ 1 $            & 1.1 & 2-25   & 7         \\
BLOSM      &  ?-         & $ 1 $           & 0.1 & 0.2-15 & 1         \\
\hline
\end{tabular}
\label{zand.table}
\end{center}
\end{table*}

\begin{table*}
\caption{Rough estimates of the masses of BLOSM key components.}
\begin{center}
\begin{tabular}{|lr|} 
\hline\hline
Component  &   Mass (kg)    \\
\hline

Optics     &    130     \\
Detectors  &    85      \\ 
Electronics&    20      \\
Spacecraft  &    125    \\
\hline
Total      &    360     \\
\hline
\end{tabular}
\label{mass.table}
\end{center}
\end{table*}

\begin{table*}
\caption{Effective areas, background count rates, and source count rates
	 in various energy bands.}
\begin{center}
\begin{tabular}{lccccc} 
\hline
Source     &  0.1-0.5 keV &  0.5-2 keV & 2-10 keV \\ 
           &   (cps)      &   (cps)    &  (cps)   \\
\hline\hline
Background      &  7.3      & 5.0        & 1.5      \\
Crab Nebula     &  1.2      & 123        & 62       \\
GRB ($n_H=0$)   &  225      & 860        & 772      \\
GRB ($n_H$=5E22) &  0        & 17         & 423      \\
XRB ($kT= 1keV$)& 0         & 0.06       & 1.2      \\
XRB ($kT= 2keV$)& 0         & 0.21       & 11.5     \\
XRB ($kT= 3keV$)& 0         & 0.37       & 30.8     \\
AGN (1 mCrab)   & 0         & 0.1        &  0.06    \\
\hline
\end{tabular}
\label{rate.table}
\end{center}
\end{table*}

\end{document}